# Problems When Studying Holography

# Probleme beim Erlernen der Holographie

Martin E. Horn, Helmut F. Mikelskis

University of Potsdam, Physics Education Research Group,
Am Neuen Palais 10, 14469 Potsdam, Germany
E-Mail: marhorn@rz.uni-potsdam.de – mikelskis@rz.uni-potsdam.de

**Abstract**

Holography as a subject is neglected largely in current physics lessons and in school books. Even though this topic might be complex, it is applicable and viable in the world that we live in. Holography lends itself to further develop knowledge with regard to inference optics. However, misconceptions of students impede the learning process.

Research conducted by our research group shows that problematic perceptions regarding interference optics still exist with physics students at the intermediate level and on occasion are even sustained by illustrations in school books. Based on this analysis, a concept for a series of lessons is introduced. This series can be taught in physics lessons at the secondary school level II and in basic courses. Initial results from dealing with the subject of holography in an advanced training course will be presented.

**Kurzfassung**

Das Thema Holographie wird im derzeitigen Physikunterricht und in Schulbüchern weitgehend vernachlässigt. Dabei handelt es sich um ein anwendungsorientiertes, in der Lebenswelt durchaus bedeutsames, wenn auch komplexes Thema. Die Holographie bietet sich an um Wissen zur Interferenzoptik weiterzuentwickeln. Allerdings hindern Fehlvorstellungen den Lernprozess.

Wie Untersuchungen unserer Arbeitsgruppe zeigen sind die problematischen Vorstellungen zur Interferenzoptik selbst bei Physikstudenten mittlerer Semester vorhanden und werden teilweise gar durch die Darstellung in Lehrbüchern gefördert. Ausgehend von dieser Analyse wird das Konzept einer Unterrichtsreihe, die in Kursen der Sekundarstufe II und im Grundstudium eingesetzt werden kann, vorgestellt. Erste Erfahrungen bei der Behandlung der Holographie im Fortgeschrittenenpraktikum werden berichtet.

Even though holography is topic geared toward application and is suitable for teaching at the upper level, it is widely neglected in school curricula. Holography, however, can be a valuable tool to further develop and expand knowledge on interference optics. The problems that arise when dealing with the topic of holography were analyzed in several preliminary studies (see Figure 1) by our research group and were taken into consideration during the conceptualization of a series of lessons on holography. The evaluation of this series was carried out at a number of Brandenburg secondary schools in the 2000/2001 school year.

Obwohl das Thema der Holographie ein anwendungsorientiertes und für die Oberstufe geeignetes Thema ist, wird es im Schulunterricht weitgehend vernachlässigt. Dabei bietet sich die Holographie an, um Wissen zur Interferenzoptik weiterzuentwickeln und zu vertiefen. Die Probleme, die bei der Behandlung der Holographie existieren, wurden in mehreren Voruntersuchungen (siehe Fig. 1) unserer Arbeitsgruppe analysiert und bei der Erarbeitung einer Unterrichtsreihe zur Holographie berücksichtigt. Die Evaluation dieser Unterrichtsreihe wird im Schuljahr 2000/2001 an mehreren Brandenburger Gymnasien durchgeführt

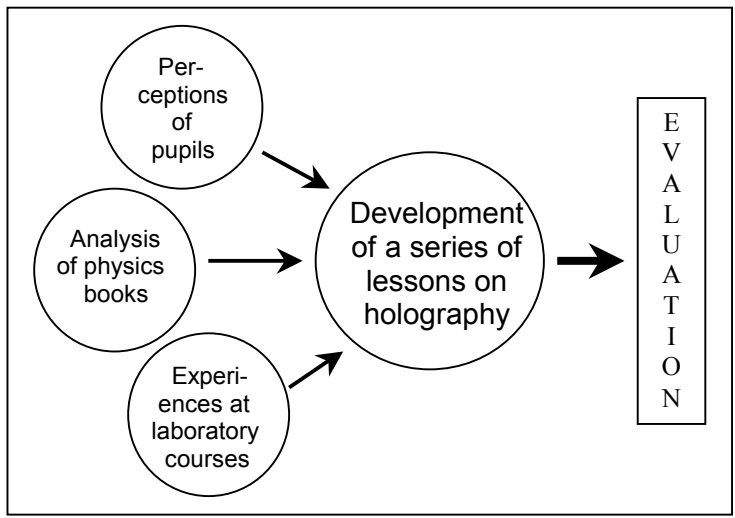

*Figure 1: Outline of the project.*

Young people already have diffuse ideas of the concept of a hologram before visiting a class on interference optics. In a questionnaire 166 students of a Berlin comprehensive school (*Gesamtschule*) were asked to impart their everyday conceptions of holography. This study was conduced in the 8[th] and 10[th] grades. Of the questioned students, allegedly slightly more than 12 percent claim to have already seen a real hologram. They characterize holograms in their descriptions by means of spatial impressions and by a change in picture as the angle of vision shifts. Here it becomes apparent that students often have great difficulties in verbalizing spatial impresssions. Students avoid these difficulties by using acronyms such as "3D." Great uncertainties also existed in distinguishing between the actual hologram and the holographic picture (see Figure 2).

Almost twice as many of the questioned students (approx. 22 %) related the term hologram to scenes in a movie. These alleged holograms induce a series of problematic perceptions. In addition to creating something from nothing via the computer (see Figure 3), the most noticeable student descriptions characterize the alleged holographic picture as a handling personality. Moreover, they sometimes assigned material properties to these figures. The depicted three-dimensional impression is however similar in all descriptions.

A small number of students related other effects with the concept of the hologram, like for example lenticular images or effects in ghost trains.

Approximately two-thirds of the questioned students had no concrete conceptions of the meaning of the term "hologram"; however, they were motivated to further investigate this concept after completing the questionnaire, in particular following the discussion of alleged holograms in science fiction movies during class.

Bereits vor einem Unterricht zur Interferenzoptik verbinden Jugendliche mit dem Begriff des Hologramms konkrete Vorstellungen. 166 Schülerinnen und Schüler einer Berliner Gesamtschule wurden mit einem Fragebogen bezüglich ihrer Alltagsvorstellungen zur Holographie befragt. Diese Untersuchung wurde in den Klassenstufen 8 - 10 durchgeführt. Lediglich etwas mehr als 12 % der befragten Schülerinnen und Schüler gaben an, bereits ein echtes Hologramm gesehen zu haben. In ihren Beschreibungen charakterisierten sie Hologramme durch einen räumlichen Eindruck und durch eine Bildänderung bei Änderung des Blickwinkels. Dabei fällt auf, dass oft große Schwierigkeiten bei der Verbalisierung des räumlichen Eindrucks bestehen. Umgangen werden diese Schwierigkeiten von Schülerinnen und Schülern durch Verwendung von Abkürzungen wie "3D". Große Unklarheiten bestanden auch in der Unterscheidung zwischen dem eigentlichen Hologramm und dem holographischen Bild (siehe Fig. 2).

Fast doppelt so viele der befragten Schülerinnen und Schüler (ca. 22 %) verbanden den Begriff des Hologramms mit filmischen Szenen. Diese vermeintlichen Hologramme induzieren eine Reihe von Fehlvorstellungen. Neben einer Erzeugung aus dem Nichts per Computer (siehe Fig. 3) fallen Schülerbeschreibungen auf, die das vermeintliche holographische Bild wie eine handelnde Persönlichkeit charakterisieren. Unter anderem ordneten sie diesen Figuren teilweise materielle Eigenschaften zu. Allen Beschreibungen gleich ist jedoch der wiedergegebene dreidimensionale Eindruck.

Einige wenige Schüler verbanden mit dem Begriff des Hologramms andere Effekte, wie beispielsweise Wackelbilder oder Effekte in der Geisterbahn.

"*This is a 3D picture that when it is looked at from different angles sometimes nothing becomes visible. Only from a particular angle, that the part on the picture is depicted so as if it was real. I have one at home.*"

"*If one moves the hologram, the hologram automatically moves with it.*"

*Figure 2: Pupil descriptions of holograms.*

> "Ein Hologramm habe ich schon einmal in einem
> „Star Trek" Film gesehen. Dort gab es einen
> Holodeck. Ein Raum wo die Umgebung durch
> den Computer hergestellt wird. z.B. Man denkt
> man ist in einem Wald, aber man ist im Holo-
> deck. Oder man sieht Personen die durch das
> Computer hergestellt ist."

*"I have already seen a hologram in a 'Star Trek' movie. There was a holodeck. A room where the surroundings are created by a computer. For example one thinks that one is in the forest, but one is in the holodeck. Or one sees persons that have been created by the computer."*

***Figure 3:*** *Pupil's descriptions of a 'hologram' in a film.*

The second preliminary study concerned an analysis of schoolbooks. The analysis was based on the thesis that physics teachers were not only hesitant to include holography as teaching material because it is considered difficult and complex, but also because the illustration in physics books often seems problematic or is completely omitted.

Of the 18 examined books appropriate for school, allegedly only five of them covered holography and that partially in only very short explanations. Didactically problematic illustrations also appear in elaborate texts. Thus, ray-optical fundamentals are not considered in the sketches. Either confusing hybrid models are implemented or unmotivated shifts between the ray model and the wave model take place. Moreover, an example from the book *"Metzler Physik (2nd ed., Stuttgart 1991)"* shows a violation of holographic principles of depiction when each object point is assigned a defined point that is set on the holographic film. Characteristic is also that the waves that originate at the hologram move toward the virtual picture (see Figure 4). Similarly simple is the usual restriction to point-like objects, so that Fresnel zone plates can allegedly be observed.

The generalization toward expanded objects is not trivial as is claimed in numerous books on physics. Other schoolbooks, on the other hand, limit themselves to explaining the production of the hologram and report on the holographic reconstruction in an allegedly descriptive manner, without referring to the physical processes that it is based on.

The often simple illustrations in school- and textbooks also received a blow in the third preliminary study, which was conducted in cooperation with students of the advanced laboratory course at the University of Potsdam. Before starting the experiment, participants were asked to create a concept map concerning holography. Five terms (hologram, interference pattern, object wave, reference wave) were provided for the students, while

Ungefähr zwei Drittel der Befragten hatten keine konkreten Vorstellungen bezüglich der Bedeutung des Begriffs "Hologramm", wurden jedoch insbesondere durch die Diskussion der vermeintlichen Hologramme in Science-Fiction-Filmen im Unterrichtsgespräch nach Ausfüllen des Fragebogens motiviert, sich mit diesem Begriff auseinander zusetzen.

Die zweite Voruntersuchung umfasste eine Schulbuchanalyse. Ausgangspunkt dieser Analyse war die These, dass die Holographie von Physiklehrerinnen und Physiklehrern nicht nur deshalb als Unterrichtsstoff zurückhaltend aufgegriffen wird, weil sie als schwierig und komplex gilt, sondern auch deshalb, weil die Darstellung in Physikbüchern oft problematisch erscheint oder gänzlich unterbleibt.

Von den 18 untersuchten schulisch einsetzbaren Büchern griffen lediglich fünf die Holographie als Thema auf, und dies teilweise in nur recht kurzen Erläuterungen. Auch bei den ausführlicheren Texten fallen didaktisch problematische Darstellungen auf. So werden strahlenoptische Grundsätze in den Skizzen nicht berücksichtigt. Es finden unmotivierte Modellwechsel zwischen Strahlenmodell und Wellenmodell statt oder es werden verwirrende Hybridmodelle eingesetzt. Ein Beispiel aus dem Buch *"Metzler Physik (2. Aufl., Stuttgart 1991)"* offenbart zudem eine Verletzung holographischer Abbildungsprinzipien, wenn jedem Objektpunkt ein fest definierter Punkt auf dem holographischen Film zugeordnet wird. Bezeichnend ist auch, dass vom Hologramm ausgehende Wellen zum virtuellen Bild wandern (siehe Fig. 4). Ebenso unbedarft ist die übliche Beschränkung auf punktförmige Objekte, so dass lediglich Fresnelsche Zonenplatten betrachtet werden.

Die Verallgemeinerung hin zu ausgedehnten Objekten ist nicht, wie in zahlreichen Physikbüchern behauptet, trivial. Andere Schulbücher wiederum they themselves had to insert at least five further central terms into their concept maps (see Figure 5). The concept maps were implemented in the sense of a knowledge test, except that the restricting rules of rigid questions did not pertain. Immediately after the drawing of the proposition arrows, the propositions were to be fixed on a special piece of paper.

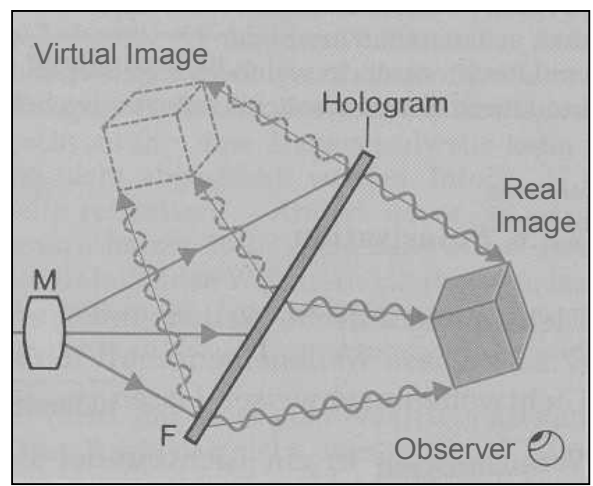

***Figure 4:*** *Formation of the virtual image according to "Metzler Physik".*

As the terms that were used in addition were tallied, it became evident that students concentrated on terms that described the structure of the experiment (e.g. laser, object, photo plate), compared to terms concerning the explanation of the physical basics of holography, which were used only occasionally. As it concerns content, the students all being in the intermediary semesters still demonstrated an extensive lack in knowledge of the appropriate vernacular as well as insecurities in the subject of physics. Thus, the network of terms illustrated in Figure 5 included formulated propositions like "*The virtual image appears as a hologram on the screen*" or "*The object sends a real image with the object wave.*"

beschränken sich auf eine Erläuterung der Hologrammaufnahme und berichten über die holographische Rekonstruktion lediglich deskriptiv ohne Rückgriff auf die ihr zugrunde liegenden physikalischen Prozesse.

Die oft unbedarften Darstellungen in Schul- und Lehrbüchern fanden ihren Niederschlag auch in der dritten Voruntersuchung, die mit Studenten im Fortgeschrittenenpraktikum an der Universität Potsdam durchgeführt wurde. Vor Versuchsbeginn sollten die Probanden jeweils ein Concept Map zur Holographie erstellen. Fünf Begriffe (Hologramm, Interferenzmuster, Kohärenzlänge, Objektwelle, Referenzwelle) wurden vorgegeben, während mindestens weitere fünf zentrale Begriffe durch die Studentinnen und Studenten selbst in ihr Begriffsnetz einzufügen waren (siehe Fig. 5). Die Concept Maps wurden damit im Sinne eines Wissenstests eingesetzt, nur dass die einschränkenden Vorgaben starrer Fragestellungen entfielen. Die Propositionen sollten sofort nach Einzeichnung des Propositionspfeils auf einem gesonderten Blatt fixiert werden.

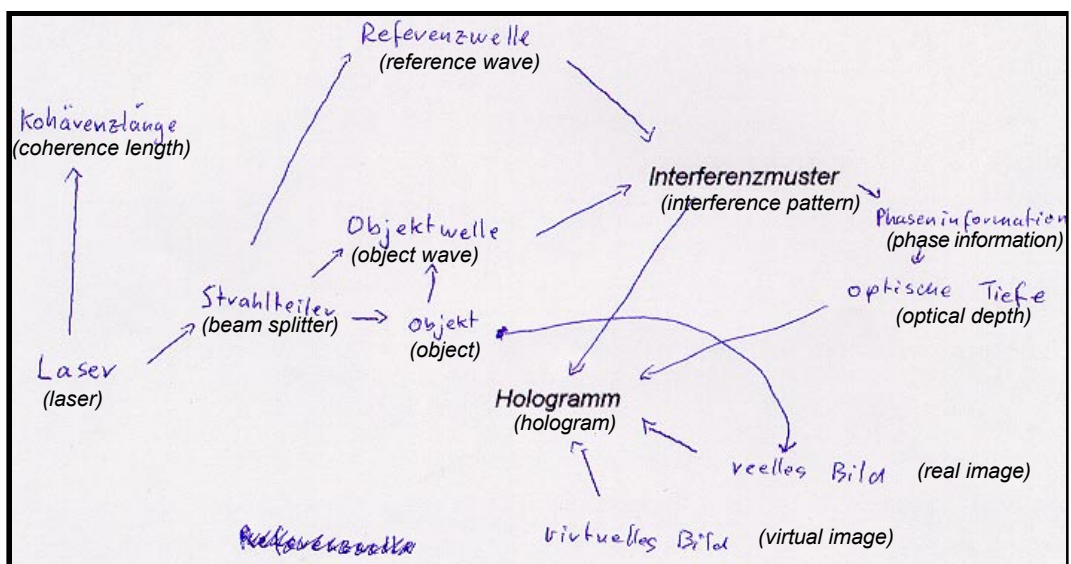

*Figure 5: Concept map of a physics student (5th semester) on holography.*

Results of the three preliminary studies flowed into the development of the series of lessons on holography. Here, the starting point is the fact that the students connect a three-dimensional structure with the concept of the hologram and they do not relate this to inference effects. The creation of simple, freely floating figures with the help of "scratch holograms" provides a motivating and functional introduction. William T. Plummer and Leo R. Gardner illustrate the physical principle of the mechanically produced preliminary stage of holography in the article "A mechanically generated hologram?" (Applied Optics, Vol. 31, No. 31/1992, p. 6585 – 6588).

Bei Auszählung der zusätzlich genannten Begriffe fällt auf, dass die Studentinnen und Studenten schwerpunktmäßig solche Begriffe hinzufügten, die den Versuchsaufbau beschreiben (z.B.: Laser, Objekt, Photoplatte), jedoch Begriffe zur Erläuterung der physikalischen Grundlagen der Holographie nur eingeschränkt verwendeten. Inhaltlich zeigten sich bei den Studenten, die sich alle in mittleren Semestern befanden, noch große fachsprachliche Mängel sowie physikalische Unsicherheiten. So fanden sich bei dem in Figur 5 gezeigten Begriffsnetz ausformulierte Propositionen wie *"Das virtuelle Bild erscheint als Hologramm auf dem Schirm"* oder *"Das Objekt sendet mit der Objektwelle ein reelles Bild"*.

A didactical reduction to reflection on concave mirrors that are carved into a piece of Perspex using circles ensures an appropriate level for students. The switch to physical holograms can be completed by discussing the interference effects on thin layers and the Fresnel zone plate. In addition to experiments what is useful in these phases are the application of Moiree models and computer simulation. Thus, it is important that the cognitive interactions of the different optical models are considered in the learning process.

Die Ergebnisse dieser drei Voruntersuchungen flossen in die Entwicklung der Unterrichtsreihe zur Holographie ein. Ausgangspunkt ist dabei die Tatsache, dass die Schülerinnen und Schüler mit dem Begriff des Hologramms eine dreidimensionale Struktur verbinden und sie diese nicht mit Interferenzeffekten in Verbindung bringen. Die Erzeugung von einfachen, frei im Raum schwebenden Figuren mittels "Ritzhologrammen" bietet dabei einen motivierenden und handlungsorientierten Einstieg. Das physikalische Prinzip dieser mecha-nisch herstellbaren Hologramm-Vorstufe wird von William T. Plummer und Leo R. Gardner in dem Artikel *"A mechanically generated hologram?"* (Applied Optics, Vol. 31, No. 31/1992, S. 6585 – 6588) dargelegt.

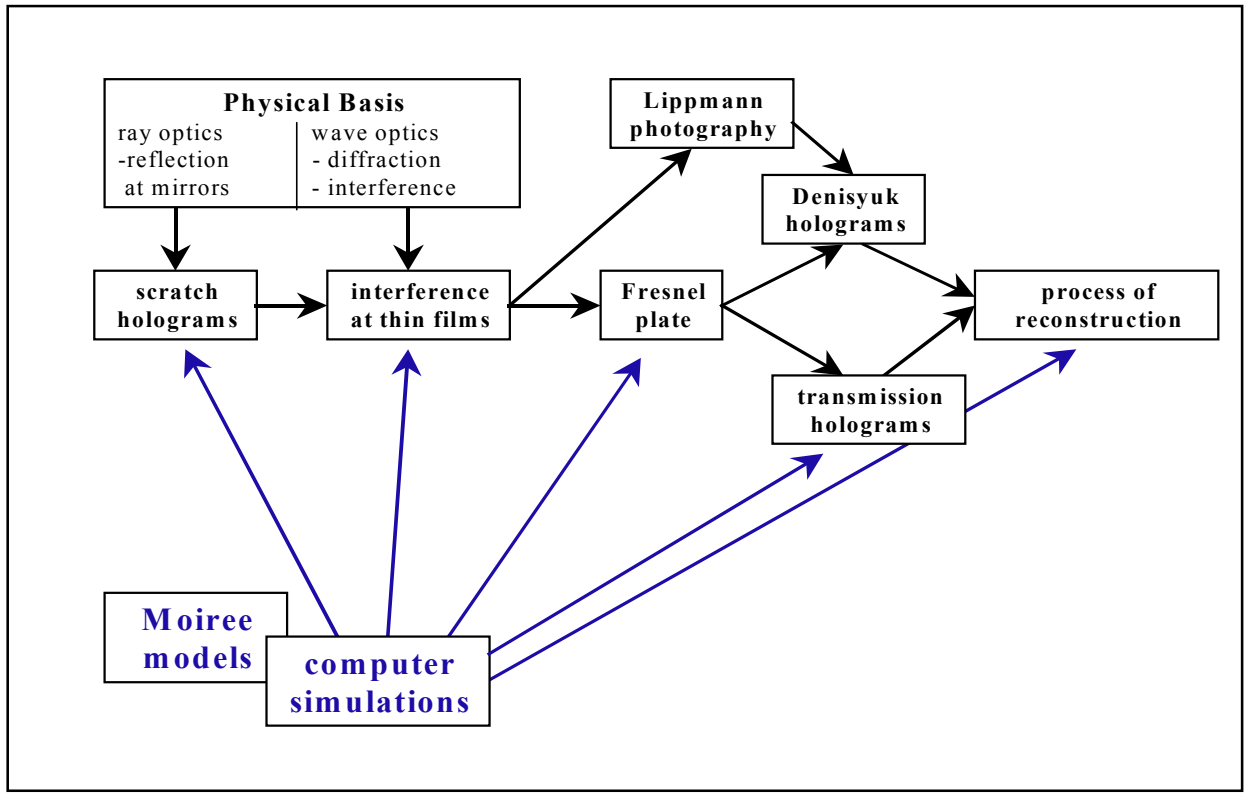

*Figure 6:* *Alternatives in structuring a series of lessons on holography.*

Eine didaktische Reduktion auf Reflexion an Hohlspiegeln, die mittels Zirkel in eine Plexiglasscheibe eingeritzt werden können, sichert ein für Schüler angemessenes Niveau. Der Übergang zu physikalischen Hologrammen kann dann über die Diskussion von Interferenzeffekten an dünnen Schichten und die Fresnelsche Zonenplatte vollzogen werden. In diesen Phasen bieten sich neben Experimenten der Einsatz von Moiree-Modellen und Computersimulationen an. Wesentlich ist dabei die Berücksichtigung kognitiver Wechselwirkungen der verschiedenen optischen Modelle im Lernprozess.